\documentclass{emulateapj}
\usepackage{apjfonts,amsmath}

\def\mbf#1{\mbox{\boldmath ${#1}$}}

\slugcomment{}

\shorttitle{Formation of Moon}
\shortauthors{Wada, Kokubo, and Makino}

\begin{document}


\title{High-Resolution Simulations of a Moon-Forming Impact and
Post-Impact Evolution}


\author{Keiichi Wada}
\affil{National Astronomical Observatory of Japan, Mitaka, Tokyo 181-8588, Japan}
\email{wada.keiichi@nao.ac.jp}
\author{Eiichiro Kokubo}
\affil{National Astronomical Observatory of Japan, Mitaka, Tokyo 181-8588, Japan}
\email{kokubo@th.nao.ac.jp}

\and

\author{Junichiro Makino}
\affil{Department of Astronomy, University of Tokyo}
\email{makino@astron.s.u-tokyo.ac.jp}




\begin{abstract}
In order to examine the ``giant impact hypothesis'' for the Moon
 formation, we run the first grid-based, high-resolution
 hydrodynamic simulations for an impact between proto-Earth and a
 proto-planet. 
The spatial resolution for the impact-generated disk is greatly improved 
from previous particle-based simulations.
This allows us to explore fine structures of a circumterrestrial debris disk
and its long-term evolution.
We find that in order to form a debris disk from which
 a lunar-sized satellite can be accumulated, the impact must result 
 in a disk of mostly liquid or solid debris, where pressure is not
 effective, well before the accumulation process starts. 
If the debris is dominated by vapor gas, strong spiral shocks are
 generated, and therefore the circumterrestrial disk cannot survive
 more than several days. 
This suggests that there could be an appropriate mass range for 
 terrestrial planets to harbor a large moon as a result of giant
 impacts, since vaporization during an impact depends on 
the impact energy.
\end{abstract}





\section{INTRODUCTION}

In the current standard scenario of planet formation, the
 final stage of assemblage of terrestrial planets is collisions among
 proto-planets of about Mars mass \citep{koku98,kort00}.
It is widely accepted that at this final assemblage stage, the Moon is
 formed from the circumterrestrial debris disk generated by an off-set impact of the
 proto-Earth with a Mars-sized proto-planet, which is known as the
 ``Giant Impact (GI) hypothesis'' \citep{hart75,came76}.
The initial phase of the GI scenario, a giant impact and the formation of
 the circumterrestrial debris disk, has been studied by a
 series of hydrodynamic simulations \citep{benz86,came00,canu01,canu04}.
N-body simulations of the accumulation of the Moon from the debris disk,
which is the last phase of GI scenario, revealed that a single
 large moon is formed just outside the Roche limit, at a distance of about
 three to four times the Earth's radius, within several months \citep{ida97,koku00}. 
It is  believed that this GI scenario explains a number of mysteries
 concerning the origin of the Moon \citep{came00}: Why is the Moon so large
 compared to satellites of other planets? 
Why is the Moon deficient in iron and volatiles compared to the Earth?
Why does the Earth-Moon system have large angular momentum? 
	
Almost all hydrodynamic simulations of GI in past decades 
 used the Smoothed Particle Hydrodynamic (SPH) method, which is a
 particle-based Lagrangian scheme \citep{lucy77,ging77}.
In the SPH method, the numerical accuracy, in other words, the
 resolution, is determined by the number of SPH particles. 
The latest simulations \citep{canu01,canu04} used
 $10^4-10^5$ SPH particles, which is an improvement of 1-2 orders
 of magnitude compared to simulations in the previous two decades \citep{benz86,came00}.
It is apparent, however, that the SPH method is not the best scheme 
 to simulate an impact between two proto-planets.
Firstly, SPH is not suitable to deal with the strong shocks and shear motion 
 produced by the off-set impact. 
Moreover, since it is basically the Lagrangian
 scheme, the current SPH does not have resolutions fine enough for
 diffuse regions.   
This is critical problem particularly for GI, because the debris disk
 consists of only a few \% of the total mass. 
As a result, even for a simulation with $10^5$ SPH particles, 
only a few $10^3$ SPH particles are used to represent the debris disk,
 and thus the fine structure of the disk is not resolved accurately.
In the SPH formalism, the spatial resolution is determined by the
 `kernel size', which is variable with the local density. 
For a diffuse circumterrestrial disk, this kernel size can be as large as
 the radius of the disk itself \citep[see Figure 1 of ][]{canu01}.  
This is the main reason that evolution of the debris is followed for
 very short period ($<1$ day) after the impact \citep{canu01}. 
Therefore the post-impact evolution of the debris, 
which is a key for the GI hypothesis, 
is not well understood.

Alternatively, one can use grid-based Eulerian methods to overcome the
 above problems.
\citet{melo89} have written the only preliminary study on the GI by the
 grid-based method to date. 
However, they neglected self-gravity of the planets and debris, and the
 evolution is followed only for less than 1 hour around the impact.
Therefore, it cannot be compared to the recent SPH simulations.

In this paper, we present the first three-dimensional hydrodynamic simulations 
of the giant impact followed by formation of a debris disk,
taking into account the self-gravity, using 
a high-accuracy Eulerian-grid scheme.
Our aim is to clarify the long-term (more than 100 hours after the
impact) evolution of the debris with the highest numerical 
accuracy used in the previous simulations for GI.
We pay attention especially to effects of pressure in the
debris on formation of the large moon, rather than exploring
the large parameter space for the mass ratio or orbits of the 
proto-planets. 
The previous SPH simulations with
$10^4-10^5$ particles would be fine for studying the behavior 
of GI for the initial several hours, and they suggest that
the fraction of mass orbiting the 
proto-Earth just after the impact is mainly determined by
the orbital parameters.
On the other hand, the long-term behavior of the debris, and therefore
the final mass of the moon depends on the evolution of the
debris disk.
It is expected that 
the pressure in the low-density debris 
should affect the hydrodynamic nature of the disk, e.g., by generating shocks.
Pressure in the debris is determined by the thermodynamics 
and the phase-change of the material after the impact, 
which is not fully understood for GI \citep{stev87}. 
For example, if a substantial fraction of the 
material is in a hot vapor form after the impact, 
the thermal pressure dominates dynamics of the disk, but
if liquid or solid is a major component of the debris,
its behavior could be very different.
Since there is no practical numerical codes to treat directly two-phase
flow and phase-change, we run
hydrodynamic simulations assuming two extreme equations of state,
which approximately represent vapor or liquid-dominated materials.

The paper is organized as follows. In \S 2, we briefly describe
our numerical method, equations of state, and initial conditions.
We show the numerical results in \S 3. We give conclusions 
and discuss an implication on a necessary condition of formation of
large satellites for the Earth-type planets in \S 4.

\section{METHOD AND MODELS}
\subsection{Numerical Methods}
The hydrodynamic scheme used here is a standard Eulerian method, which has been
widely applied to various astrophysical problems involving strong shocks 
in structures with a high density contrast \citep{wada01a, wada01b}.
\setcounter{footnote}{0} 
We solve the following conservation equations for mass, momentum,
and energy, and the Poisson equation numerically in three dimensions:
  \begin{eqnarray}
\frac{\partial \rho}{\partial t} + \nabla \cdot (\rho \mbf{v}) &=& 0,
\label{eqn: rho} \\ \frac{\partial \mbf{v}}{\partial t} + (\mbf{v}
\cdot \nabla)\mbf{v} +\frac{\nabla p}{\rho} + \nabla \Phi_{\rm sg} &=& 0, \label{eqn: rhov}\\
 \frac{\partial (\rho E)}{\partial t} +  \nabla \cdot 
[(\rho E+p)\mbf{v}] &=& 0, \label{eqn: en}\\ \nabla^2
\Phi_{\rm sg} &=& 4 \pi G \rho, \label{eqn: poi} 
\end{eqnarray}
 where, $\rho,p,\mbf{v}$, and $E$ are the density, pressure, velocity of
the matter, and the specific total energy. The gravitational potential
of the matter is denoted by $\Phi_{\rm sg}$.


We use AUSM (Advection Upstream Splitting Method) \citep{liou93} to
solve the hydrodynamic equations and MUSCL (Monotone  Upstream-centered Schemes for Conservation Laws) to achieve third-order spatial accuracy.
Gravitational potential of the fluid on grid points is calculated by solving eq. (\ref{eqn: poi}) using
 the Fast Fourier Transform with a convolution method \citep{hock81}.
These are the standard techniques for hydrodynamics taking into account
self-gravity in grid-based schemes.
More detailed descriptions on the numerical scheme and test calculations to
guarantee the numerical accuracy are described in Wada \& Norman (2001).

We use an equally spaced Cartesian grid for 
the computational box; 20$\times$20$\times$5 (model A and model B)
in units of $r_{\rm E}$, where $r_{\rm E}$ is the radius of the proto-Earth.
In order to see how the results are affected by a size of computational 
domain, we also run models in a larger domain, i.e. 40$\times$40$\times$ 10 (model A' and model B').
The total number of grid points in the computational box is
$512\times 512\times 124 \simeq 3.3 \times 10^7$ in all the models.
The circumterrestrial region (i.e. $r > r_{\rm E}$) is represented 
by $\sim 10^7$ grid points.


\subsection{Equation of State}
It is still an open question that what kind of equation of state (EOS)
 should be used in order to simulate complicated thermal and dynamical
 processes during GI and post-impact evolution. 
Conventionally two types of empirical EOS have been used with the SPH method: 
 the Tillotson EOS \citep[e.g.][]{canu01, till62, benz99} and ANEOS and its extension \citep[e.g.][]{melo00, canu04}.
ANEOS approximately describes mixed phase states (e.g., vapor and melt) by  
 treating the different phases as separate components in temperature and
 pressure equilibrium. 
Note that there are no numerical simulations for GI that can handle
 two-phase flow consistently, and that spatial resolution must be
much smaller than the system size to describe the mixed phase states.
As mentioned above, it is apparent that the spatial resolutions in
previous SPH simulations are insufficient to describe mixed phase states,
even if they use ANEOS (see also \S 3.3).

One should also note that recent SPH simulations \citep{canu01,canu04} showed that the initial behavior (i.e. a few dynamical times) of the impact
 does not strongly depend on choice of EOS for a given geometry of
 impact (i.e., mass ratio between proto-planets and the impact
 parameter).  
This suggests that the mass and the angular momentum of the debris produced by
 the impact, which are the most important parameters for the accumulation
 process of a moon, are determined by a gravitational process, not by
detailed thermal processes described by the conventional EOSs. 

We run a series of numerical experiments to follow a long-term ($>$ 20
 dynamical times) evolution of the debris disk.
Instead of using the conventional EOSs, such as ANEOS or Tillotson EOS,  
 we here assume two extreme EOSs: a polytrope-type EOS (EOS-1) 
 and the same EOS, but with a zero-pressure cut-off (EOS-2).
This is because assuming a simple EOS is more suitable to clarify essential physics
behind the numerical results.
EOS-1 is 
\begin{eqnarray}
p = (\gamma -1)\rho E + C(n/3 + 1 - \gamma)\rho^{n/3 + 1}, 
\label{eqn: eos1}
\end{eqnarray}
where $E$ is the internal energy, and $C$ is a constant. 
Throughout this paper, we assume the ratio of specific heats, $\gamma=1.01$,
and the constants $C =1$, and $n =12$.  
With EOS-1, the system behaves like an ideal gas or Polytropic gas at 
 high or low temperature limits, respectively.  EOS-2 is
the same as EOS-1, but we introduce cut-off density ($\rho_c = 2.7$ g cm$^{-3}$) and internal energy ($E_c = 4.72\times 10^{10}$ erg g$^{-1}$),
below which the pressure is forced to be zero in equations (\ref{eqn: rhov}) and
(\ref{eqn: en}).
Note that in the present problem, density in most of 
the debris is below $\rho_0$.
These numbers are the same in the SPH simulations in which
the Tillotson EOS is used \citep{benz99}.
We do not adopt the `negative pressure' regime 
assumed in the Tillotson EOS \citep{benz99}, because
it is apparently unphysical on the scale of GI. 
With this negative pressure regime, condensation
in diffuse gas on any scales, e.g. the scale of the Moon, is allowed,
even if self-gravity of the material does not work.

The essential behavior of EOS-2 is similar to that of the Tillotson EOS. 
The cut-off for pressure phenomenologically represents a transition 
from vapor to liquid or solid particles.  
On the other hand, EOS-1 represents hot gaseous debris.
The thermodynamical process during GI is complicated, for example,
 it is still unclear what fraction of the impactor is vaporized.
This can be solved by a numerical scheme which can handle a two-phase (vapor
 and liquid) flow with elementary physics, but there is no practical
 code for the GI problem at this moment.
We therefore perform numerical experiments for GI using hydrodynamic
 simulations with these two extreme EOSs to clarify the effect of pressure
 on post-impact evolution.
\subsection{Initial Conditions}
%
We follow \citet{canu01} and \citet{canu04} for the orbital parameters of
the impactor for which the most massive satellite is expected.
The masses of the proto-Earth 
 and the impactor are assumed to be $1.0M_\oplus$ and $0.2M_\oplus$,
where $M_\oplus$ is the Earth mass.
The radii of the proto-Earth and proto-planet 
are $r_E = 1.0$ and 0.64$r_E$, respectively.
Note that no significant differences in the results
for smaller impactors (e.g. $0.1M_\oplus$) were found in our simulations.  
The initial orbits of the impactor are assumed to be parabolic, and the
 angular momentum is 0.86 $L_{\rm graz}$, where $L_{\rm graz}$ is the
 angular momentum for a grazing collision \citep{canu01}.
Initially the impactor is located at 4.0$r_{\rm E}$ from the
 proto-Earth.  

\section{RESULTS}
\subsection{Disk Evolution and the Predicted Lunar Mass}
Figure 1 shows a typical time evolution of the giant impact with EOS-1 
 (model A).
This model corresponds to the `late' impact model in \cite{canu01}.
After the first impact ($t \simeq 1 $ hour), the disrupted impactor is
 reaccumulated to form a clump at $t \simeq 3$ hours, which finally
 collides with the proto-Earth at $t \simeq 6$ hours.  
During the second impact, the impactor is destroyed, and 
 a dense part of the remnant spirals onto the proto-Earth ($t \simeq 10$
 hours), and a circumterrestrial debris disk is formed around 
 $t \simeq 18$ hours. 
It should be noted that many strong spiral shocks are generated in this
 process as seen in the density map (Figure 2) and azimuthal density
profile (Figure 3). 

Figure 4 is the same model as model A (Figure 1), but with EOS-2 (model
 B). 
The initial behavior is similar to that of  model A.
The impactor is completely destroyed after the second impact as is the
 case in model A.
However, the remnant is not diffuse, but more condensed (Fig. 4)
 and geometrically thin (see the edge-on views).
This difference is reasonable, because with EOS-2, the internal pressure
 does not work in the low density gas. 

Next, we predict the satellite mass accumulated from the circumterrestrial
 disk in our models.  
Equation (\ref{eqn: lunarmass}) is an empirical formula for the satellite mass ($M_{\rm s}$) as
 a function of the specific angular momentum of the circumterrestrial
 disk ($j_{\rm disk}$) derived from $N$-body experiments of lunar accretion \citep{koku00}:
\begin{eqnarray}
 \frac{M_{\rm s}}{M_{\rm disk}} \simeq 1.9 
  \frac{j_{\rm disk}}{\sqrt{GM_\oplus a_{\rm R}}}
  - 1.1 -1.9 \frac{M_{\rm esc}}{M_{\rm disk}}, 
\label{eqn: lunarmass}
\end{eqnarray}
 where $M_{\rm disk}$ is the disk mass, $a_{\rm R}$ is the Roche limit radius,
 and $M_{\rm esc}$ is escaped mass, which is the mass lost during the
 accumulation process.  
Using eq.~(\ref{eqn: lunarmass}) and time-dependent mass and angular momentum taken from
 our numerical results as initial conditions for the accumulation process, 
 we plot the time evolution of the {\it predicted} lunar mass in Figure 5.
Here  $M_{\rm esc}/M_{\rm disk}$ is assumed to be 0.05 \citep{koku00, canu04}.
As seen in Fig. 5, the predicted lunar mass reaches a maximum of 
 $\simeq 1.4M_L$ for model A and  $\simeq 1.2 M_L$ for
 model B just after the second impact ($t\simeq 10$ hours). 
However, it should be noted that the predicted mass for the model A 
 decreases very rapidly after the lunar mass reaches the maximum on a
 time-scale of 10 hours.   
The predicted lunar mass becomes smaller than the current lunar mass
 $M_L$ after $t\simeq 15$ hours, and almost monotonically
 declines to $0.3M_L$ at $t \simeq 65$ hours. 
On the other hand, if we use EOS-2 which is relevant for a mostly liquid or
solid material, the lunar mass increases
 again after $t \simeq 40$ hours (model B).
It reaches to $\simeq 0.6 M_L$, and stays nearly constant until
 $t \simeq$ 90 hours.

In a fixed-grid Euler method, a choice of the spatial resolution and size of 
the computational box is a trade-off.
If the computational volume is not large enough, 
the material that escapes from the computational domain 
in the initial phase of the impact
cannot be followed, 
and this could reduce the final mass of the circumterrestrial disk. 
In order to check this effect, we also run model B', in which
the computational volume  ($40\times 40\times 10 r_E^3$) is eight-times
larger than that in model B (i.e. the grid size is twice larger than
that in model B).
The behavior is qualitatively similar with the result in model B, but
the maximum of the predicted mass is about 1.5 time larger (2$M_L$)
than that in model B. 
This is because that a part of remnant `arc' of the impactor is lost
from the computational box in model B in the initial phase of the impact 
(see the snapshots at $t=14.2$ hours in Fig. 3).
As a result, the predicted lunar mass in a quasi-steady state ($t > 60$ hours)
is also 50\% larger than that in model B. 
However, model B' also shows the lunar mass does not change after $t\simeq 50$ hours, supporting the conclusion that the long-term evolution of the debris disk is followed
correctly by the numerical resolution of model B'.
Since the computational volume in model B' is large enough to cover 
more than 90\% mass of the remnant arc formed by the impact, 
we expect that the predicted lunar mass cannot be much larger than
0.9$M_L$, even if we use a much larger computational volume
with the same spatial resolution in model B.
Similarly, model A' covers 8 times larger volume than model A, and the 
peak lunar mass is 1.6 times larger than the one in model A. However, 
it also shows rapid decrease of the lunar mass, which is in contrast 
with the behavior in model B and B'.


\subsection{Angular Momentum Transfer by Spiral Shocks}
The reason why the predicted lunar mass cannot stay nearly constant
 after the impact in model A is clear. 
As seen in Figures 2 and 3, spiral shocks are generated in the
 debris disk.  
Figure 3 shows that at least two clear shocks are
present for the same radii (i.e. $\phi = 0.2 \pi$ and 1.95 $\pi$ 
for $r = 2.9r_E$ and $\phi = 0.5\pi$ and 1.5$\pi$ for  $r= 5.0r_E$). The density jump at each shock is a factor of 4 to 16, suggesting that
the effective Mach number in the model is about 2 to 4.
In fact, a significant fraction of the debris is supersonic in model A.
Figure 6, which is a frequency distribution of Mach number $\cal{M}$ in the debris at two
different epochs, clearly shows that $\cal{M}$ $> 6$ (mass-weighted)
or $\cal{M}$ $> 3$ (volume-weighted) in majority of the debris 
at $t=12.3$ hours. Even in a later stage, i.e. $t=57.1$ hours, the disk is still supersonic $\cal{M}$ $\gtrsim 2$. 


Due to these shocks in the gaseous disk, the inner massive disk inside the Roche limit
 effectively falls onto the proto-Earth in a few rotational periods (see discussion
below).    
This initial fall of the disk cannot be avoided in the GI process, 
if majority of the debris is in a hot gas phase.
Since the collision should be off-set, 
the generation of the spiral shocks is inevitable in a vapor gas 
 disk with a  high Mach number. 

The angular momentum transfer by the spiral shocks is more effective for 
stronger shocks with larger pitch angles (i.e., more open-spirals).
 We simply estimate it as 
follows. For isothermal shocks, the post-shock velocity is $\sim v_0/{\cal M}^2$,
where $v_0$ is pre-shock velocity of the 
gas, perpendicular to the shock front.  Therefore the stream line
is bent by passing an oblique shock \citep[see Fig. 12 and 
Appendix in][]{wada04}. In a rotating medium, this effect 
removes the angular momentum
of a fluid element when it passes a standing spiral shock. 
The angular momentum change due to one passage of a spiral shock 
can be approximately estimated by 
\begin{eqnarray}
j/j_0 \approx \left( \frac{\sin^2 i}{{\cal M}^4} + \cos^2 i \right)^{1/2}
\cos\left[i - \arctan \left(\frac{\sin i}{{\cal M}^2 \cos i}\right) \right],
\label{eqn: ang}
\end{eqnarray}
where $i$ is the pitch angle of the spiral shock, $j_0$ and $j$ are
the angular momentum of pre- and post-shock flow \citep{wada04}.
The sound velocity of SiO$_2$ gas with 2000 K is $0.6 $ km s$^{-1}$ 
for 100\% vapor, and $\simeq 0.1$ km s$^{-1}$ if the mass ratio
between vapor and liquid is 0.1 \citep{stev87}.
For strong shocks (${\cal M} \gg 1$),
the angular momentum change, equation (\ref{eqn: ang}), is $j/j_0 \sim \cos^2 i$.
 Therefore, the angular momentum loss is
$\simeq $ 12\% per spiral shock with $i = 20^\circ$. 
For weak shocks (e.g., ${\cal M} = 2$), equation (\ref{eqn: ang}) suggests that 
the angular momentum loss per shock is about 5.6\% for $i = 20^\circ$ and 
1.4\% for $i = 10^\circ$.
As shown in Fig. 2 and Fig. 3, several
spiral shocks are generated in the debris disk, then a fluid element would 
lose $\sim$ 10\% of its angular momentum for one rotational period. 
This is large enough for the debris gas to lose most of their
angular momentum in 100 hours after the impact. 
Therefore, the presence of spiral shocks in the debris disk is crucial
for evolution of the disk, and as a result for the mass of satellites.

\subsection{Comparison with Previous SPH Results}
Remarkably, although we use methods, resolutions, and equations of state
 different from previous SPH simulations, the predicted mass just
after the impact is consistent with recent SPH results: $1.7-1.8 M_L$ for the `late impact'  model \citep{canu01,canu04}. 
This again implies that the early phase of the impact process ($t \lesssim 10$
 hours) is dominated by gravity, but not by hydrodynamical and thermal 
 processes.  
However, we find that behavior of the remnant and 
the predicted lunar mass in the late phase  
depend on EOS, especially whether pressure works in the disk.
This was not pointed out by the previous
SPH simulations. One obvious reason is that the spatial resolutions
in the SPH results are not fine
enough to follow long-term evolution of the debris
until the effect of pressure becomes evident.
In the previous SPH
simulations \citep{canu04}, 
only $\sim 10^3$ particles are used to represent the debris.
Suppose 1000 SPH particles are uniformly 
distributed in a thin disk inside the Roche limit ($3r_E$),
 the average separation between particles is  $\simeq 0.2 r_E$.
In SPH, variable kernels are often used, 
and the kernel size is roughly the same as the spatial resolution.
Usually an SPH particle is assumed to interact with 30-50 neighbor particles. 
Therefore, the kernel size in the previous SPH simulations, 
may be $\sim r_E$ in the disk, which is about 10-20 times
larger than our grid size. Moreover, the debris has in fact
an extended distribution, rather than a thin disk, and the SPH kernel size is
much larger than the numbers estimated above.
It is not straightforward to compare the spatial resolution
between the two different schemes, but apparently that the spatial 
resolution in our simulation is greatly improved from the previous ones.
For example, in Figure 3, the azimuthal density profile is obtained
by about 1000-2000 grid points. 
However, in the previous SPH simulations, only 10-100 particles would be
available for drawing this kind of density profile. 

The other reason why the previous studies did not notice 
the effect of shocks is that 
the conventional EOSs (ANEOS and Tillotson) are
close to our EOS-2, in a sense that pressure is not effective in the
diffuse debris.


%
\section{CONCLUSIONS AND DISCUSSION}
We run for the first time three-dimensional grid-base hydrodynamic 
simulations of the giant impact between proto-Earth and a proto-planet
 taking into
account self-gravity. We assume two types of the equation of state
that phenomenologically represent hot gaseous material 
or liquid/solid material.
Our numerical experiments suggest that in order to form a
 lunar-sized satellite from the circumterrestrial debris disk produced
 by the giant impact, the most fraction of the debris should not be in
a pressure-dominated phase (e.g., hot vapor gas).
Otherwise the subsequent disk evolution results in forming only a small
 satellite because of the fast angular momentum transfer associated with
spiral shocks prior to the satellite accretion stage.
The time-scale of the angular momentum transfer is an order of 10 days,
therefore the accumulation to form a large satellite 
should be much faster than this.
Yet our results may not rule out a possibility forming a satellite as large
as the Moon from a pressure-dominated disk, if the disk mass is much
larger than the current lunar mass.
This might be the case for 
collisions with a massive impactor, but this causes another problem 
on a fraction of vapor in the debris (see discussion below).

In agreement with the previous SPH simulations \citep[e.g.,][]{canu01, canu04},
we find that the predicted lunar mass at $t\simeq 10$ hours after 
the collision does not strongly depend on
choice of EOS and numerical methods for the same orbital parameters.
This means that the early phase of 
GI is dominated by a gravitational process, not by thermodynamical 
processes. On the other hand, the late phase of GI, i.e.,
evolution of the debris disk, is sensitive to EOS, especially 
the pressure in the disk. In order to clarify this difference, 
the spatial resolutions of the previous SPH simulations were apparently 
not fine enough.

Our results give an important implication on a necessary condition for formation
 of large satellites for the Earth-type planets. 
The present results suggest that one of the important keys for GI
 scenario is the fraction of the vaporized debris after the giant
 impact. 
If the kinetic energy of the collision is much larger than the latent
 heat of the major component of the proto-planet, 
 most of the proto-planet could be vaporized by
 the impact. 
If this is the case, the disk evolution would not lead to formation of a
 large moon as explained above.  
On the other hand, if the impact velocity is slower than a critical
 value, a large fraction of the material can be in a liquid or solid
 phase, and shocks do not dominate the angular momentum transfer in the
 debris disk. 
In this case, the post-impact evolution of the proto-planet could be
 similar to that in model B in our experiments, and therefore a single
 large moon could be formed. 
By assuming the impact is head-on and comparing the latent heat of
 SiO$_2$ \citep{stev87} and the kinetic energy of the impact, we can roughly
 estimate the critical impact velocity as $\simeq $15 km s$^{-1}$, which
 is slightly larger than the surface escape velocity of the Earth.
In the grazing impact between proto-Earth and a Mars-sized proto-planet,
 shocks propagating in the planets involve vaporization,
and its velocity is probably comparable (or
 smaller) to this critical value, and therefore it would be natural to
 postulate that most of the debris mass is in a liquid phase.
This argument leads to an interesting suggestion: if the impact velocity
 is larger than the critical velocity, in other words, the mass of the
 planet is larger than, say, a few Earth mass, the giant impact never
 results in forming a large satellite.

Finally, one should recall that a correct phase-change even for the
 major component, such as SiO$_2$, during GI followed by formation
 of a debris disk is still unknown \citep[e.g.][]{stev87}. 
For example, the post-impact expansion could produce more vapor in the
 debris material, and this may change the hydrodynamic property of the
 circumterrestrial disk.  
The final fate, namely whether the planet has satellites, and how large
 they are, depends on the nature of the phase-change.
The thermo-dynamical process during catastrophic impact between
 the proto-planets is still too complicated to be explored using
 current numerical techniques including the SPH and grid-based methods.
This ultimately requires a self-consistent numerical scheme to simulate
 two-phase flow taking into account the realistic thermodynamical
 processes.   

\acknowledgments

We are grateful to Yutaka Abe for his fruitful comments.
Numerical computations were carried out on Fujitsu VPP5000 at NAOJ.  
The authors also thank the anonymous referee for his/her valuable comments 
and suggestions.

\clearpage

\begin{figure}
\includegraphics[height=11cm]{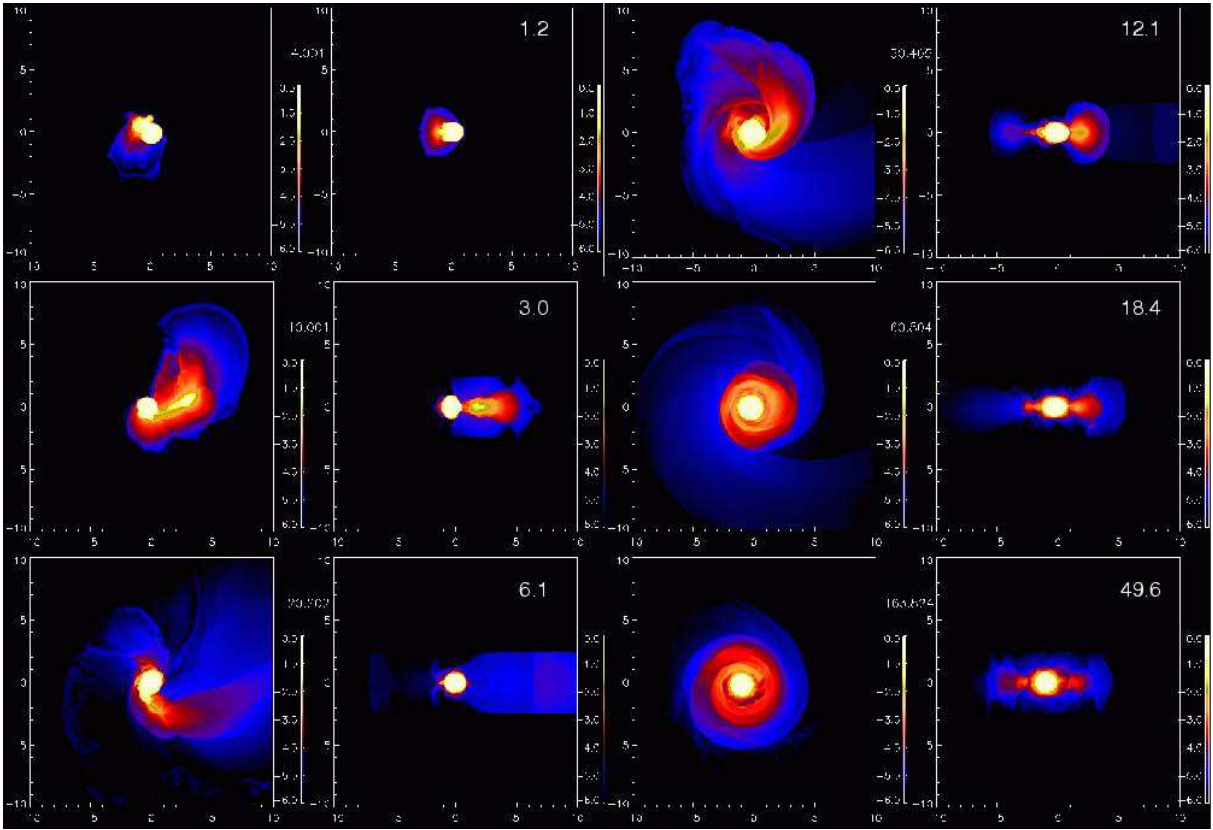}
\caption{
Giant impact  simulation with EOS-1, which 
represents a state where most of the impactor mass is 
vaporized.
The panels in the left and right rows are face-on and edge-on views
 of the system, respectively.
The figures in the panels show the time in units of hours.
The color represents log-scaled density (The unit is $\rho_0 = 12.6$ g cm$^{-3}$). 
}
\end{figure}


\begin{figure}
\includegraphics[height=7cm]{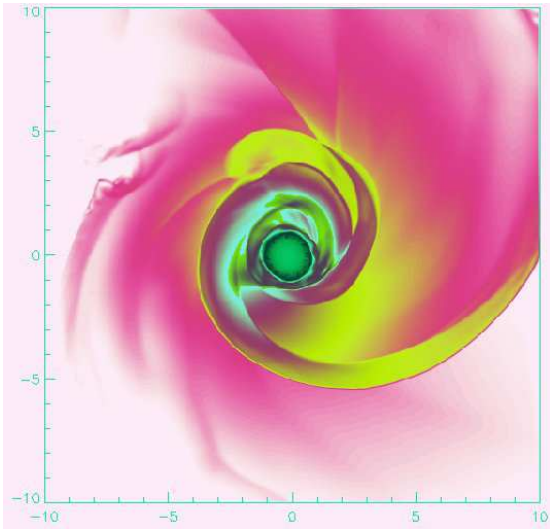}
\caption{
Snapshot of the density field of model A at $t=12.3$ hours.
Strong spiral shocks in the debris are resolved. 
}
\end{figure}


\begin{figure}
\includegraphics[height=7cm]{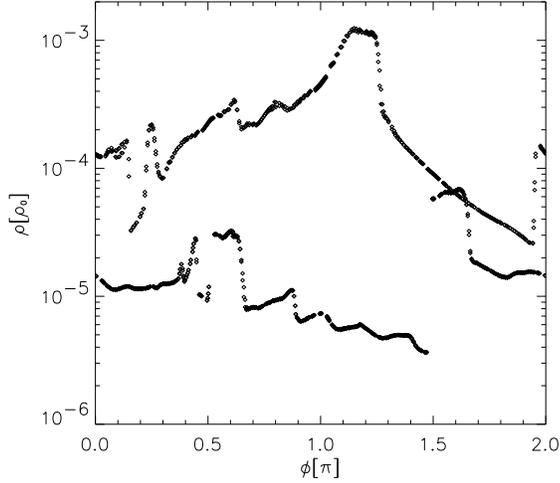}
\caption{
The azimuthal density profiles at $r=$ 2.9 $r_E$ (upper dots) and 5.0 $r_E$ (lower dots) on $z=0$ plane for the same density snapshot in  
Fig. 2. The unit of density is $\rho_0 = 12.6$ g cm$^{-3}$.
}
\end{figure}


\begin{figure}
\includegraphics[height=12cm]{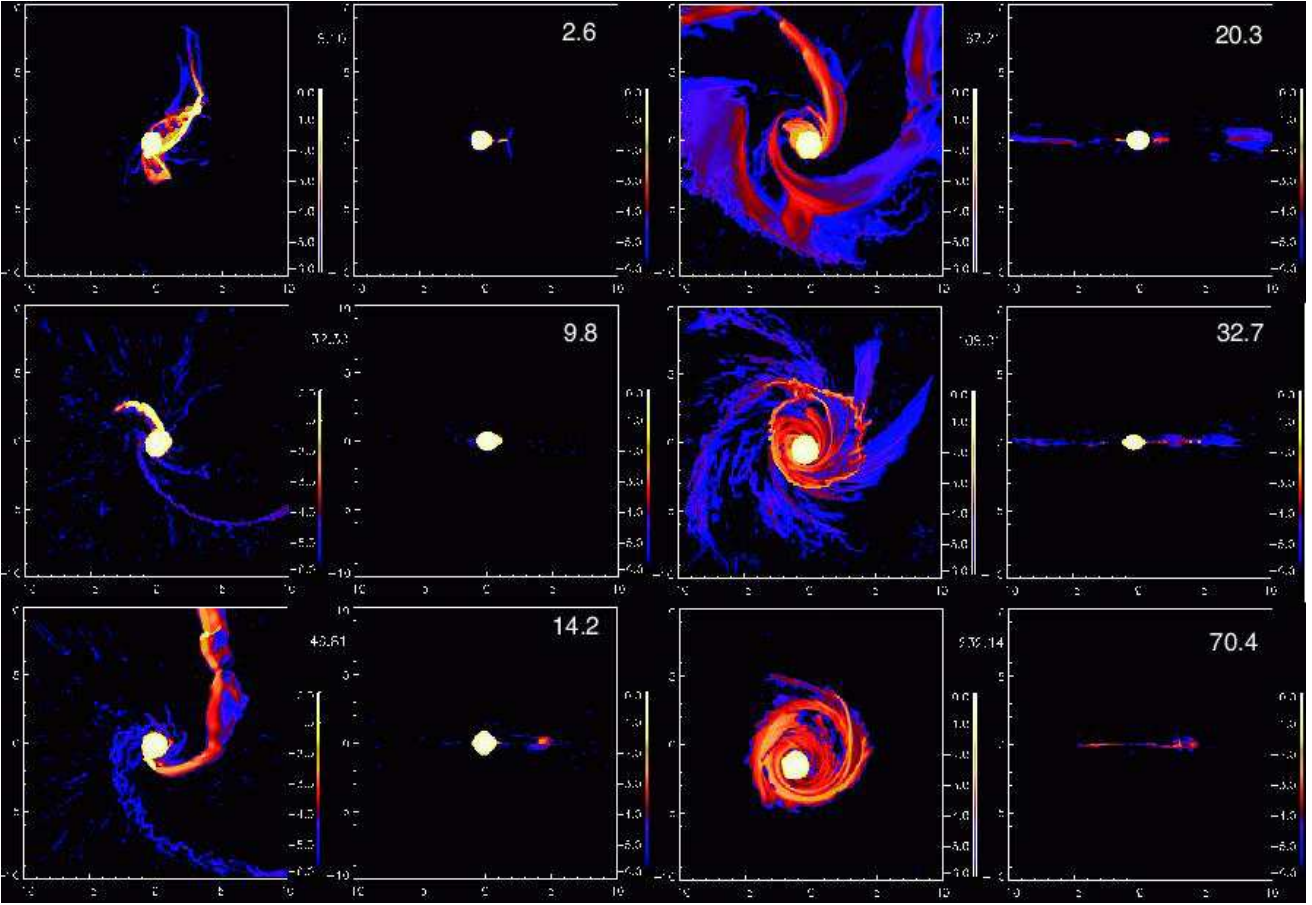}
\caption{
Same as in Figure 1, but for model B, in which
zero-pressure EOS (EOS-2) is assumed.
Following previous simulations with the Tillotson EOS \citep{till62}, 
we adopt the cut-off density ($\rho_c = 2.7$ g cm$^{-3}$) and 
internal energy ($E_c = 4.72\times 10^{10}$ erg g$^{-1}$), below 
which the pressure is forced to be zero in EOS-1.
The EOS-2 represents a state where most of the mass is in a state of
liquid, not in vapor.}
\end{figure}


\begin{figure}
\includegraphics[height=10cm]{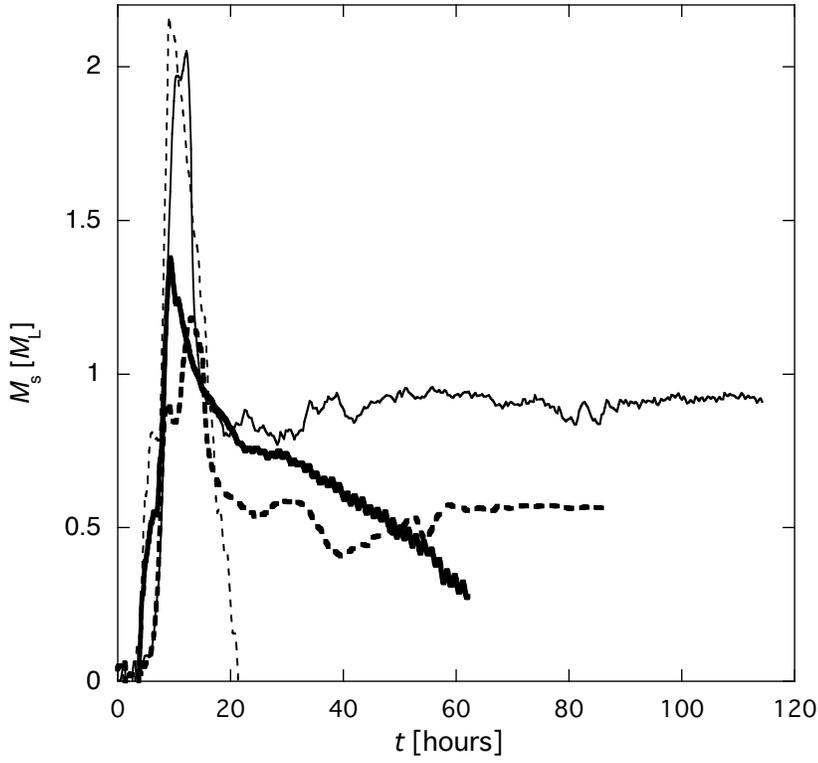}
\caption{
Evolution of the predicted lunar mass ($M_{\rm s}$) in model A 
 shown in Figure 1 (thick solid line) and model B in Figure 4 (thick 
dashed line). 
The vertical axis is normalized by the current lunar mass ($M_L = 0.0123 M_\oplus$).
The thin dashed and solid lines show models A' and B', in which 
the computational volumes are eight-times larger
than those in models A and B.
}
\end{figure}


\begin{figure}
\includegraphics[height=10cm]{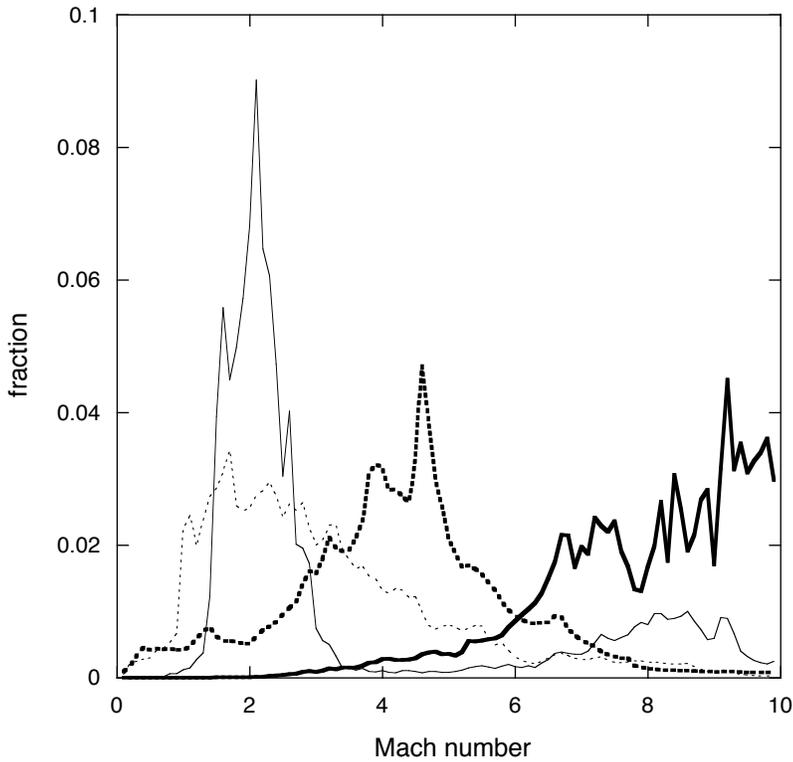}
\caption{
Frequency distribution of Mach number 
in the debris ($\rho < 0.1\rho_0$) of model A.
Thick  lines are at $t = 12.3$ hours.
Thin  lines are at $t = 57.1$ hours.
Solid and dotted lines are mass-weighted and volume-weighted histograms, 
respectively.
}
\end{figure}

\end{document}